# Coarse-Grained Model of the Demixing of DNA and Non-Binding Globular Macromolecules

Marc Joyeux*

*LIPHY, Université Grenoble Alpes and CNRS, Grenoble, France*

(*) email : marc.joyeux@univ-grenoble-alpes.fr

**ABSTRACT:** The volume occupied by the unconstrained genomic DNA of prokaryotes in saline solutions is thousand times larger than the cell. Moreover, it is not separated from the rest of the cell by a membrane. Nevertheless, it occupies only a small fraction of the cell called the nucleoid. The mechanisms leading to such compaction are the matter of ongoing debates. The present work aims at exploring a newly proposed mechanism, according to which the formation of the nucleoid would result from the demixing of the DNA and non-binding globular macromolecules of the cytoplasm, like ribosomes. To this end, a coarse-grained model of prokaryotic cells was developed and demixing was analyzed as a function of the size and number of crowders. The model suggests that compaction of the DNA is actually governed by the volume occupancy ratio of the crowders and remains weak almost up to the jamming critical density. Strong compaction is however observed just before jamming, suggesting that crowding and electrostatic repulsion work synergetically in this limit. Finally, simulations performed with crowders with different sizes indicate that the DNA and the largest crowders demix preferentially. Together with the recent observation of the gradual compaction of long DNA molecules upon increase of the concentration of BSA proteins and silica nanoparticles, this work supports the demixing mechanism as a key player for the formation of the nucleoid.

**INTRODUCTION**

The genomic DNA of eukaryotes is separated from the cellular cytoplasm by a nuclear envelope and displays several levels of compaction ranging from the initial wrapping of the DNA helix around histone proteins to the final X-shape of chromosomes. In contrast, the genomic DNA of prokaryotes lacks such a detailed organization and is not separated from the rest of the cell by any membrane. As has been known for decades, it nevertheless occupies only a fraction of the cell called the *nucleoid*, which is rather surprising because the volume occupied by the unconstrained molecule in saline solutions is several thousands of times larger than the volume of the cell. The mechanism leading to the compaction of the bacterial genomic DNA has puzzled the scientists for decades[1] and is still the matter of ongoing debates.[2] The crucial point is that none of the mechanisms proposed until recently provide a convincing explanation for the formation of the nucleoid. More precisely, supercoiling provokes only mild compaction[2,3] and the number of nucleoid associated proteins capable of bridging two DNA duplexes is too small to induce significant global compaction.[2,4] In contrast, the conjunction of DNA charge neutralization by small polycations[5] and the action of fluctuation correlation forces[6] is able to compact the DNA significantly, but this is essentially an all-or-none mechanism, with the DNA molecule being either in the coil state or in a globular state much denser than the bacterial nucleoid.[7,8] Such an abrupt transition from the coil state to a too dense globule is also observed upon addition to the buffer of long neutral polymers and salt[9] (to compensate for the weakness of depletion forces[10]) or long anionic polymers.[11,12] Finally, long cationic polymers are able *in vitro* to compact progressively the DNA molecule to shrunken coil structures that resemble those of the DNA inside the nucleoid,[8] but this associative phase separation mechanism cannot play a role in the compaction of the bacterial DNA *in vivo*, because prokaryotes cells do not contain sufficient amounts of long polycations or proteins with large positive charges.

However, it has been shown recently that negatively charged globular macromolecules may also play a role in the formation of the nucleoid. Indeed, long DNA molecules could be compacted gradually to densities comparable to that of the nucleoid by adding 5 to 10% (w/v) of bovine serum albumin (BSA) to the solution[13,14]. It can be argued that this result is not completely unambiguous, because the surface of BSA proteins displays small positively charged patches despite its total charge of approximately -18*e* and the formation of weak BSA-DNA coacervates has been reported.[15] Fortunately, it has been checked even more recently that a few percents of negatively charged silica nanoparticles with diameters ranging from 20 to 135 nm are also able to compact gradually the DNA,[16] thereby confirming the results obtained with BSA. These results suggest that the compaction of the bacterial nucleoid may result from the demixing of DNA and other non-binding globular macromolecules contained in the cytoplasm,[17] this hypothesis being all the more sensible as approximately 30% of the dry mass of cells is composed of ribosomes, which are almost spherical and highly negatively charged complexes with diameter 20-25 nm. The formation of the nucleoid would consequently result from a segregative phase separation[18] leading to a phase rich in DNA (the nucleoid) and another phase rich in the other macromolecules (the rest of the cytoplasm). This hypothesis has received little attention up to now, although it has been evoked on theoretical grounds almost 20 years ago.[19,20]

The purpose of the present work is to elaborate further on this putative mechanism for nucleoid compaction by addressing two important questions. The first one deals with the determination of the volume occupancy ratio at which segregation takes place. It is indeed not easy to estimate the volume effectively occupied by proteins or nanoparticles in a given experiment, because the Debye length cannot be determined precisely. While calculations suggest that silica nanoparticles occupy approximately 15-20% of the volume at the point of DNA collapse,[16] the authors nevertheless suspect that the actual Debye length was actually larger than the value they used in their calculations and speculate that '*the observed critical concentrations of nanoparticles approach the overlapping concentration, which is ca. 74% for most densely packed spheres*'[16]. The second question relates to size dispersion of macromolecules in the cytoplasm. Indeed, experiments are usually performed with macromolecules and/or particles that are all of the same size, while the size of macromolecules in the cytoplasm varies over a large range. It is consequently important to get an idea of how size dispersion affects demixing in real conditions. It should be stressed that these two questions are rather difficult to handle from a purely theoretical point of view, because the outcome of statistical physics calculations depends critically on the quality of the description of the interactions between crowders and/or between crowders and the DNA chain.[21,22] In the present work, we therefore sought for answers through simulations based on a coarse-grained model that will now be described.

**COMPUTATIONAL DETAILS**

As illustrated in Fig. 1, the coarse-grained model used in this work consists of a circular chain of $n = 1440$ beads of radius $a = 1.78$ nm separated at equilibrium by a distance $l_0 = 5.0$ nm (the genomic DNA) enclosed in a large confining sphere of radius $R_0 = 120$ nm (the cell), together with $N$ spheres of radius $b$ (the crowding macromolecules). As in previous work,[23-28] each bead represents 15 consecutive DNA base pairs, so that the model may be thought of as a 1:200 reduction of a typical *E. coli* cell with a nucleotide concentration around 10mM, close to the physiological value. The potential energy of the system, $E_{pot}$, is the sum of four terms

$$E_{pot} = V_{DNA} + V_{DNA/C} + V_{C/C} + V_{wall} , \qquad (1)$$

which describe the internal energy of the DNA molecule, the DNA-crowder interactions, the crowder-crowder interactions, and the repulsive potentials that maintain the DNA and the crowders inside the confining sphere, respectively. The internal energy of the DNA molecule, $V_{DNA}$, is further written as the sum of 3 contributions[23-28]

$$V_{DNA} = \frac{h}{2} \sum_{k=1}^{n} (l_k - l_0)^2 + \frac{g}{2} \sum_{k=1}^{n} \theta_k^2 + e_{DNA}^2 \sum_{k=1}^{n-2} \sum_{K=k+2}^{n} H(\|\mathbf{r}_k - \mathbf{r}_K\|) , \qquad (2)$$

where

$$H(r) = \frac{1}{4\pi\varepsilon r} \exp\left(-\frac{r}{r_D}\right) , \qquad (3)$$

which describe the stretching, bending, and electrostatic energy of the DNA chain, respectively. $\mathbf{r}_k$ denotes the position of DNA bead $k$, $l_k$ the distance between two successive beads, and $\theta_k$ the angle formed by three successive beads. The stretching energy is a computational device without biological meaning, which is aimed at avoiding a rigid rod description. $h$ was set to $1000\, k_B T / l_0^2$, in order for $|l_k - l_0|$ to remain on average as small as $0.02\, l_0$ despite the forces exerted by the crowders (in this work, all energies are expressed in units of $k_B T$, where $T = 298$ K). In contrast, the bending rigidity constant, $g = 9.82\, k_B T$, was chosen so as to provide the correct persistence length for DNA, $\xi = g l_0 / (k_B T) \approx 49$ nm, equivalent to 10 beads.[29] Finally, the electrostatic energy is expressed as a sum of Debye-Hückel potentials[30], where $e_{DNA} = -12.15\, \bar{e}$ (with $\bar{e}$ the absolute charge of the electron) denotes the value of the charge placed at the centre of each DNA bead,[23] $\varepsilon = 80\, \varepsilon_0$ the dielectric constant of the medium, and $r_D = 1.07$ nm the Debye length inside the medium. Since each bead represents 15 consecutive DNA base pairs, the total charge carried by the corresponding phosphate groups is $-30\, \bar{e}$, but the smaller value $e_{DNA} = -12.15\, \bar{e}$ was used in the Debye-Hückel potential to take counter-ion condensation into account.[30] Moreover, the small value of the Debye length, $r_D = 1.07$ nm, corresponds to a concentration of monovalent salts close to 100 mM and reflects the order of magnitude of the Debye length expected in bacterial cells.[16] Admittedly, the equilibrium separation of two DNA beads, $l_0 = 5.0$ nm, is too large compared to the value of the Debye length to warrant that different parts of the DNA chain will never cross. However, such crossings are very infrequent and appear to affect the geometry of the DNA chain only in a limited fashion. Therefore, the term of the potential energy that would fully prevent chain crossing (Eq. (A.5) of Ref. 2), which is quite expensive from the computational point of view, was discarded from the lengthy simulations reported here. Finally, electrostatic interactions between nearest-neighbours are not included in Eq. (2), because it is considered that they are already accounted for in the stretching and bending terms.

In the same spirit, the DNA-crowder and crowder-crowder interactions are expressed as sums of Debye-Hückel potentials with hard cores

$$V_{\text{DNA/C}} = e_{\text{DNA}} \, e_C \sum_{k=1}^{n} \sum_{j=1}^{N} H(\|\mathbf{r}_k - \mathbf{R}_j\| - b)$$

$$V_{\text{C/C}} = e_C^2 \sum_{j=1}^{N-1} \sum_{J=j+1}^{N} H(\|\mathbf{R}_j - \mathbf{R}_J\| - 2b) \,, \quad (4)$$

where $H(r)$ is the function described in Eq. (3), $e_C$ is an electrostatic charge that characterizes the spherical crowders and $\mathbf{R}_j$ denotes the position of crowding sphere $j$. The $e_C^2 H(r-2b)$ potential in Eq. (4) differs from the usual DLVO potential[31,32]

$$W^{\text{DLVO}}(r) = \frac{e_M^2}{4\pi\varepsilon r (1+\frac{b}{r_D})^2} \exp(-\frac{r-2b}{r_D}) \,, \quad (5)$$

where $e_M$ is the total charge of the sphere, essentially by the fact that it diverges for a separation $r$ between the centers of the spheres equal to $2b$, while the DLVO potential diverges for $r = 0$. However, for

$$e_M = e_C (1 + \frac{b}{r_D}) \sqrt{\frac{r_0}{r_0 - 2b}} \tag{6}$$

the $e_C^2 H(r - 2b)$ and $W^{DLVO}(r)$ potentials remain close to each other in a broad range of separation values around $r_0$. For example, Fig. 2 shows that the $e_C^2 H(r - 2b)$ potential with $e_C = e_{DNA} = -12.15 \, \bar{e}$ and $b = 6.5$ nm remains close to the DLVO potential with $e_M = -210 \, \bar{e}$ for all interaction values ranging from 0 to $8 \, k_B T$. In this work, we choose to use the $e_C^2 H(r - 2b)$ potential rather than the $W^{DLVO}(r)$ potential, because the increase in the effective width of the crowder due to electrostatic interactions does not depend on the radius $b$ of the spheres for the former one, while it decreases with increasing $b$ for the later one. As will be emphasized in the Results and Discussion Section, this feature makes the interpretation of the results obtained with the $e_C^2 H(r - 2b)$ potential particularly straightforward. The same argument holds of course for the $e_{DNA} e_C H(r - b)$ potential that describes DNA-crowder interactions. The three electrostatic repulsion potentials, $e_{DNA}^2 H(r)$, $e_{DNA} e_C H(r - b)$, and $e_C^2 H(r - 2b)$, are plotted in Fig. 3 for $e_C = e_{DNA}$ and $b = 6.5$ nm. Most simulations discussed in the present paper were performed with $e_C = e_{DNA}$, but a series of simulations were performed with $e_C = 0.8 e_{DNA}$ for the sake of comparison. Finally, $V_{wall}$ is written in the form

$$V_{wall} = \zeta \, (\sum_{k=1}^{n} f(\|\mathbf{r}_k\|) + \sum_{j=1}^{N} f(\|\mathbf{R}_j\|)) \, , \tag{7}$$

where the repulsive force constant $\zeta$ was set to $1000 \, k_B T$ and the function $f(r)$ is defined according to if $r \leq R_0$ : $f(r) = 0$

if $r > R_0$ : $f(r) = \left(\dfrac{r}{R_0}\right)^6 - 1$ . (8)

The dynamics of the system was investigated by integrating numerically overdamped Langevin equations. Practically, the updated positions at time step $n+1$ were computed from the positions at time step $n$ according to

$$\mathbf{r}_k^{(n+1)} = \mathbf{r}_k^{(n)} + \dfrac{\Delta t}{6\pi \eta a} \mathbf{f}_k^{(n)} + \sqrt{\dfrac{2 k_B T \Delta t}{6 \pi \eta a}} \, x_k^{(n)}$$
$$\mathbf{R}_j^{(n+1)} = \mathbf{R}_j^{(n)} + \dfrac{\Delta t}{6\pi \eta b} \mathbf{F}_j^{(n)} + \sqrt{\dfrac{2 k_B T \Delta t}{6 \pi \eta b}} \, X_j^{(n)} \, ,$$ (9)

where $\Delta t = 20$ ps is the integration time step, $\mathbf{f}_k^{(n)}$ and $\mathbf{F}_j^{(n)}$ are vectors of inter-particle forces arising from the potential energy $E_{pot}$, $T = 298$ K is the temperature of the system, $x_k^{(n)}$ and $X_j^{(n)}$ are vectors of random numbers extracted from a Gaussian distribution of mean 0 and variance 1, and $\eta = 0.00089$ Pa s is the viscosity of the buffer at 298 K. After each integration step, the position of the centre of the sphere was slightly adjusted so as to coincide with the centre of mass of the DNA molecule.

**RESULTS AND DISCUSSION**

Simulations were performed by first letting the DNA chain equilibrate inside the confining sphere. Since the unconstrained chain forms a coil with a radius of gyration larger than $\sqrt{n l_0 \xi / 6} \approx 242$ nm (this estimation based on the Worm-Like Chain model neglects electrostatic repulsion between DNA segments),[33] which is substantially larger than the radius $R_0 = 120$ nm of the confining sphere, the relaxed chain occupies the whole space inside the sphere and the repulsive potential $V_{wall}$ acts directly on some of the DNA beads to repel them inside the confining sphere, thereby preventing further expansion of the chain. A typical conformation of the equilibrated chain is shown in the bottom left vignette of Fig. 1. Simulations indicate that the mean value of the radius of gyration of the equilibrated confined chain is close to 82 nm for a Debye length $r_D = 1.07$ nm. A number *N* of crowding spheres with radius *b* are then introduced at random (but non-overlapping) positions inside the confining sphere, as shown in the top vignette of Fig. 1, and the system is allowed to equilibrate again. After equilibration times ranging from a few ms (at low volume occupancy) to more than 10 ms (close to jamming), the system reaches a new steady state, which is characterized by a (usually) lower value of the radius of gyration of the DNA chain, as shown in the bottom right vignette of Fig. 1. Practically, simulations were performed with four different values of *N* (*N*=500, 1000, 2000, or 3000 crowding spheres) and 8 to 15 different values of *b* for each value of *N*. The results are shown in Fig. 4, where the mean value of the radius of gyration of the DNA chain after equilibration with the crowding spheres, $R_g$, is plotted as a function of the radius *b* of the crowders. It is observed in this figure that the four curves corresponding to the different values of *N* display the same evolution as a function of *b*, which can be divided into three different regimes. Indeed, for the lowest values of *b*, $R_g$ decreases regularly down to a threshold value close to $R_g \approx 62$ nm. Close examination of the simulations reveals that, over the whole range of corresponding values of *b*, the repulsive potential $V_{wall}$ still contributes to repelling directly some of the DNA beads inside the confining sphere. In contrast, for the largest values of *b*, the system is jammed and neither the DNA chain nor the crowders are able to move significantly. As a result, $R_g$ remains close to its initial value of 82 nm. Finally, in a rather narrow interval of values of *b* located between these two regimes, $R_g$ drops sharply to values as low as about 50nm. Strikingly, in this regime no part of the DNA chain is any longer in contact with the confining sphere, so that $V_{\text{wall}}$ contributes uniquely to maintaining the crowders inside the confining sphere, while compaction of the DNA chain results exclusively from interactions among DNA beads and crowding spheres. It is certainly not by chance that, for the four values of $N$, maximum compaction of the DNA chain occurs just before jamming. The obvious interpretation of this observation is that crowding conditions close to the jamming transition constrain DNA and the crowders to remain at short distances from one another and feel electrostatic interactions despite the shortness of the Debye length, which ultimately favors demixing of DNA and the crowders and compaction of the DNA chain.

Examination of Fig. 4 further indicates that the optimum value of $b$ leading to maximum DNA compaction decreases with increasing $N$, being close to 11.5, 8.5, 6.5, and 5.5 nm, for $N$=500, 1000, 2000, and 3000, respectively. Estimation of the volume occupancy ratio according to $N(b/R_0)^3$ suggests that the optimum ratio also decreases with increasing $N$, being close to 0.44, 0.36, 0.32, and 0.29, for $N$=500, 1000, 2000, and 3000, respectively. However, this estimation neglects the electrostatic repulsion between crowders. A natural way of taking such repulsion into account consists in considering that the effective radius $b+\Delta b$ of a crowding sphere is equal to half the separation at which the interaction energy of two crowders is equal to the thermal energy $k_B T$. The interesting point in using the $e_C^2 H(r-2b)$ potential of Eq. (4) instead of the DLVO potential of Eq. (5) to describe the interaction between two crowders is precisely that $\Delta b$ does not depend on $b$ for the $e_C^2 H(r-2b)$ potential and a particular value of $e_C$, while it does depend on $b$ for the DLVO potential and a particular value of $e_M$. As illustrated in Fig. 3, $\Delta b$ is close to 1.8 nm for $e_C = e_{\text{DNA}}$ and $r_D = 1.07$ nm. It is quite noteworthy that estimation of the effective volume occupancy ratio according to

$$\rho = N(\frac{b+\Delta b}{R_0})^3 \qquad (10)$$

leads to nearly identical values of $\rho$ at maximum DNA compaction for all values of $N$, namely $\rho \approx 0.68$, 0.63, 0.66, and 0.68, for $N$=500, 1000, 2000, and 3000, respectively. Even more striking is the fact that the four plots for $R_g$ collapse on a single master curve when $\rho$ is chosen as the abscissa axis instead of $b$, as can be seen in Fig. 5. This plot indicates that, within the validity of the model, the compaction dynamics of the DNA chain is driven exclusively by the effective volume occupancy ratio of the crowders. More precisely, the radius of gyration of the DNA chain decreases linearly with increasing values of $\rho$ up to $\rho \approx 0.6$, while significantly stronger DNA compaction, which no longer requires the direct action of the confining wall on the DNA beads, takes place in the range $0.6 \leq \rho \leq 0.7$. Finally, the system becomes jammed around $\rho \approx 0.75$, meaning that jamming occurs for the soft spheres considered here for values of $\rho$ only slightly larger than for hard spheres ($\rho \approx 0.66$).[34]

The leading role of $\rho$ is further confirmed by the plot of the mean value of the interaction energy between the confining wall and the DNA chain as a function of $0.75 - \rho$, the gap to the volume occupancy ratio at jamming, which is shown in Fig. 6. Again, the curves corresponding to the four different values of $N$ collapse on a single master curve. Moreover, the use of log-log axes highlights the fact that the average work exerted by the wall to maintain the DNA chain inside the confining sphere decreases as the cube of $0.75 - \rho$ in the whole range of values of $\rho$ extending from weak crowding to the threshold where strong compaction takes place. While this power law is probably model-dependent, this graph nevertheless emphasizes the fact that $\rho$ controls not only the profile of DNA concentration inside the confining sphere but also the strength of the interactions between the DNA chain and the confining wall.

Finally, let us mention that the results presented above do not depend on the exact value of $e_C$, the parameter that characterizes the electrostatic charge carried by the crowding sphere. To ascertain this point, a set of simulations were run with $N$=2000 crowders and $e_C = 0.8 e_{DNA}$, that is for crowder/crowder interactions reduced by a factor 0.64 compared to simulations run with $e_C = e_{DNA}$. Still, it proved to be sufficient to plug the appropriate values of $\Delta b$ in Eq. (10), that is $\Delta b = 1.6$ nm for $e_C = 0.8 e_{DNA}$ against $\Delta b = 1.8$ nm for $e_C = e_{DNA}$, for the plots obtained with the two different values of $e_C$ to superpose (results not shown).

The simulations reported above are therefore consistent with the progressive compaction of long DNA molecules, which is observed upon increase of the concentration of anionic silica nanoparticles added to the solution containing the DNA molecule.[16] They furthermore support the conjecture of the authors that maximum compaction is observed when the '*concentrations of nanoparticles approach the overlapping concentration*',[16] which is the first point this work aimed to ascertain. The simulations moreover indicate clearly that, close to jamming, crowding and electrostatic repulsion forces work synergetically in favor of particularly efficient DNA-crowders demixing and strong compaction of the DNA chain.

The second point addressed in this work deals with the effect of crowders size dispersion, and more precisely how macromolecules size dispersion may affect demixing in real cells. To investigate this point, several sets of simulations were run with crowders with two different radii enclosed simultaneously in the confining sphere and compared with simulations involving only mono-disperse crowders. All sets of simulations with bi-disperse crowders actually led to similar results and the discussion below focuses on the comparison of results obtained with 1400 spheres with radius $b$=7.2 nm and 600 spheres with radius $b$=3.5 nm, on one side, and 2000 crowders with radius $b$=6.5 nm, on the other side. In both cases, the effective volume occupancy ratio is close to the optimum value for maximal DNA compaction ($\rho \approx 0.65$). Practically, we will compare the density distributions $p_X(r)$, such that the mean number of particles of species X, which centers are located in a distance interval $[r, r+dr[$ from the center of the confining sphere, is $4\pi r^2 n_X p_X(r) dr$, where $n_X$ is the total number of particles of type X ($n_X = n = 1440$ for DNA, $n_X = N = 2000$ for 2000 crowders with radius $b$=6.5 nm, *etc*). Fig. 7A shows the distribution obtained after equilibration of 2000 crowders with radius $b$=6.5 nm in the absence of DNA. The regularly spaced peaks, which are separated by about 14.5 nm, denote the quasi-crystalline order that prevails inside the confining sphere just below the jamming critical density. The distribution becomes progressively flatter with decreasing values of $\rho$. Similarly, the distributions obtained after inserting the 2000 crowders with radius $b$=6.5 nm at random positions in the confining sphere containing the pre-equilibrated DNA chain and allowing the system to equilibrate again, as for the results in Figs. 4-6, are shown in Fig. 7B. As discussed above, the DNA beads and the crowding spheres demix, with the effect that the concentration of DNA beads increases steeply close to the center of the sphere but vanishes beyond 90 nm, while the concentration of crowders decreases regularly towards the center of the confining sphere compared to the system without DNA (Fig. 7A). These distributions can be contrasted with those shown in Figs. 7C and 7D, which were obtained along the same lines, but with 1400 spheres with radius $b$=7.2 nm and 600 spheres with radius $b$=3.5 nm instead of 2000 spheres with radius $b$=6.5 nm. Quite interestingly, it may be observed that, in the absence of DNA (Fig. 7C), the outer shell is composed almost exclusively of spheres with the largest radius, while the two distributions remain roughly equal for all the inner shells. Note also that the crystalline order is progressively lost when moving away from the edge of the sphere. Finally, Fig. 7D shows the distributions obtained with both the DNA chain and bi-disperse crowders. It is clearly seen in this figure that the distribution of smaller crowders varies little compared to the system without DNA (Fig. 7C), while the distribution of larger crowders decreases regularly towards the center of the confining sphere, as for mono-disperse crowders (Fig. 7B). Stated in other words, the total volume fraction occupied by crowders of any size is the crucial quantity for strong DNA compaction, while segregation out of the nucleoid affects primarily the crowders with largest size. All in all, these results suggest that, in real prokaryotic cells, demixing involves the DNA molecule and the largest globular macromolecules present in the cytoplasm, that is essentially the ribosomes, although their concentration is smaller than the critical crowder concentration leading to strong compaction.

**CONCLUSIONS**

Triggered by the experimental observation of the gradual compaction of long DNA molecules upon increase of the concentration of BSA proteins[13,14] and silica nanoparticles[16], the main purpose of the present work was to elaborate further on the conjecture that the formation of the bacterial nucleoid may be driven by the demixing of DNA and non-binding globular macromolecules present in the cytoplasm. For this purpose, several sets of simulations were performed by varying the number and size of spherical crowders in a coarse-grained model of prokaryotic cells. Demixing of DNA and the crowders and compaction of the DNA chain were observed in all simulations starting from homogenously distributed particles. The simulations moreover highlight the fact that the gradual compaction of the DNA chain is governed by the volume occupancy ratio of the crowders. DNA compaction increases with this ratio but remains weak almost up to the jamming critical density. Much stronger DNA compaction is however observed just before jamming, suggesting that crowding and electrostatic repulsive interactions work synergetically in favor of particularly efficient demixing. Finally, simulations performed with crowders of different sizes indicate that the DNA chain and the largest crowders demix preferentially.

It should furthermore be stressed that, in the simulations discussed above, the volume fraction occupied by naked crowders at the critical concentration for strong DNA compaction ranges from 0.29 to 0.44, depending on the number $N$ of crowders, so that the volume fraction left for virtual water ranges from 0.56 to 0.71, which is in qualitative agreement with the 50-70% water content that is usually reported for prokaryotic cells. Moreover, the translational diffusion coefficient of macromolecules is known to be much smaller in prokaryotic cells than in water and in eukaryotic cells,[35] which indicates that the cytoplasm of prokaryotic cells is indeed close to jamming. Put together, these two facts confirm that the mechanism described here may indeed play an important role *in vivo*.

This work therefore adds weight to the hypothesis that the formation of the nucleoid in bacteria may actually result from the demixing of DNA and non-binding globular macromolecules present in the cytoplasm, that is, in other words, from a segregative phase separation[18] leading to a phase rich in DNA (the nucleoid) and another phase rich in the other macromolecules (the rest of the cytoplasm).[13,14,16] Since about 30% of the dry mass of prokaryotic cells is composed of ribosomes, which are highly and almost uniformly charged anionic complexes with diameter 20-25 nm, and are excluded from the nucleoid in their functional form,[36,37] it is furthermore tempting to argue that ribosomes (and complexes made of ribosomes) actually play the role of the larger crowders in the coarse-grained model, while most other non-binding macromolecules play the role of the smaller crowders, which are only weakly segregated or not segregated at all.

To conclude, let me argue that it is probably confusing to describe the demixing mechanism discussed above as resulting from the action of 'depletion forces', in spite of the fact that it is governed by the volume occupancy ratio of the crowders. Indeed, the term

'depletion force' traditionally describes the effective attraction force between macromolecules, which results from the preferential exclusion of smaller cosolutes from the vicinity of these macromolecules. Such depletion forces can be dominated either by entropy[10], as is probably the case for the condensation of DNA by neutral polymers[9], or by enthalpy,[38,39] as may be the case for the condensation of DNA by anionic polymers.[11,12] In both cases, depletion forces are very short-ranged and compact the DNA molecule very abruptly to almost crystalline densities above a certain polymer concentration threshold. When the diameter of the crowders is much larger than the diameter of the DNA duplex, as is the case here, it is instead more natural to compare the strength of the pair interaction between DNA and the crowder, on one side, with the average strength of the pair interactions between two DNA segments and between two crowders, on the other side, to get an insight whether the system will remain globally homogenous or will demix. This difference in the strength of pair interactions is precisely the $\chi$ parameter of Flory-Huggins polymer solution theory,[18] which sign determines whether the two species demix ($\chi > 0$) or not ($\chi < 0$), and which absolute value determines the extent of demixing, which is consequently gradual instead of being an all-or-none process, as the condensation provoked by depletion forces.

## ACKNOWLEDGEMENT

This work was supported by the Centre National de la Recherche Scientifique and Université Grenoble Alpes.

**Initial** (crowders shown); $R_g \approx 82$ nm

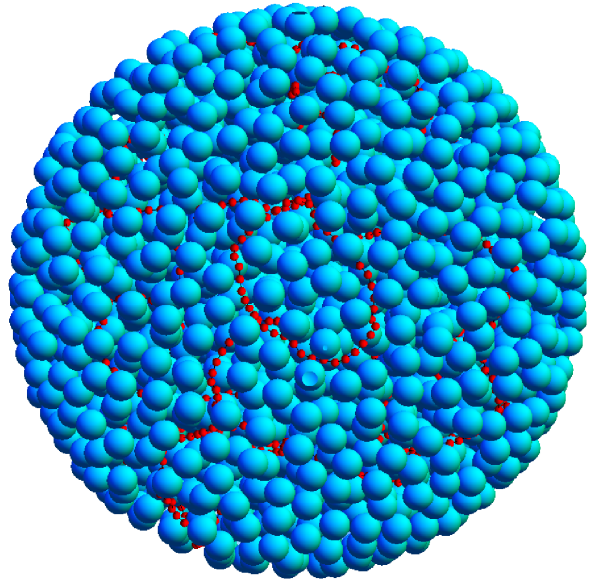

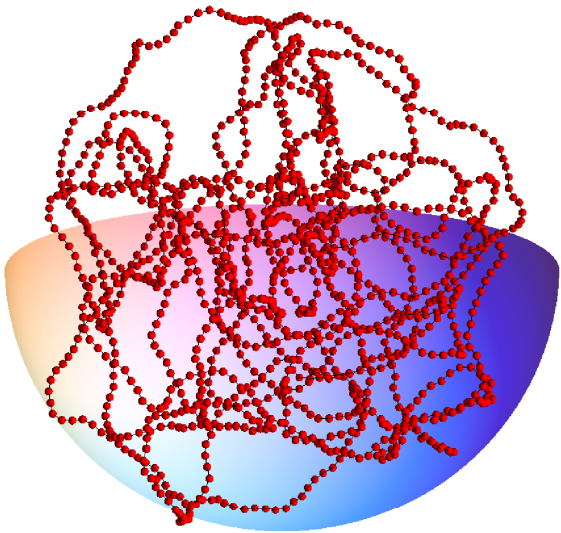
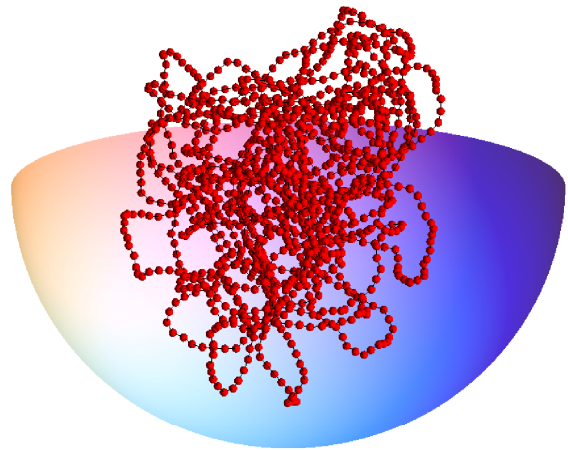

**Initial** (crowders not shown)
$R_g \approx 82$ nm

**Final** (crowders not shown)
$R_g \approx 50$ nm

**Figure 1** : The top vignette shows a typical conformation of the equilibrated DNA chain with 1440 beads (red) and the 2000 crowding spheres with radius $b = 6.5$ nm (cyan), which have just been added randomly. The bottom left vignette shows the same snapshot with the crowders having been removed and ¼ of the confining sphere being shown. The bottom right vignette shows a typical conformation of the DNA after equilibration with the crowding spheres.

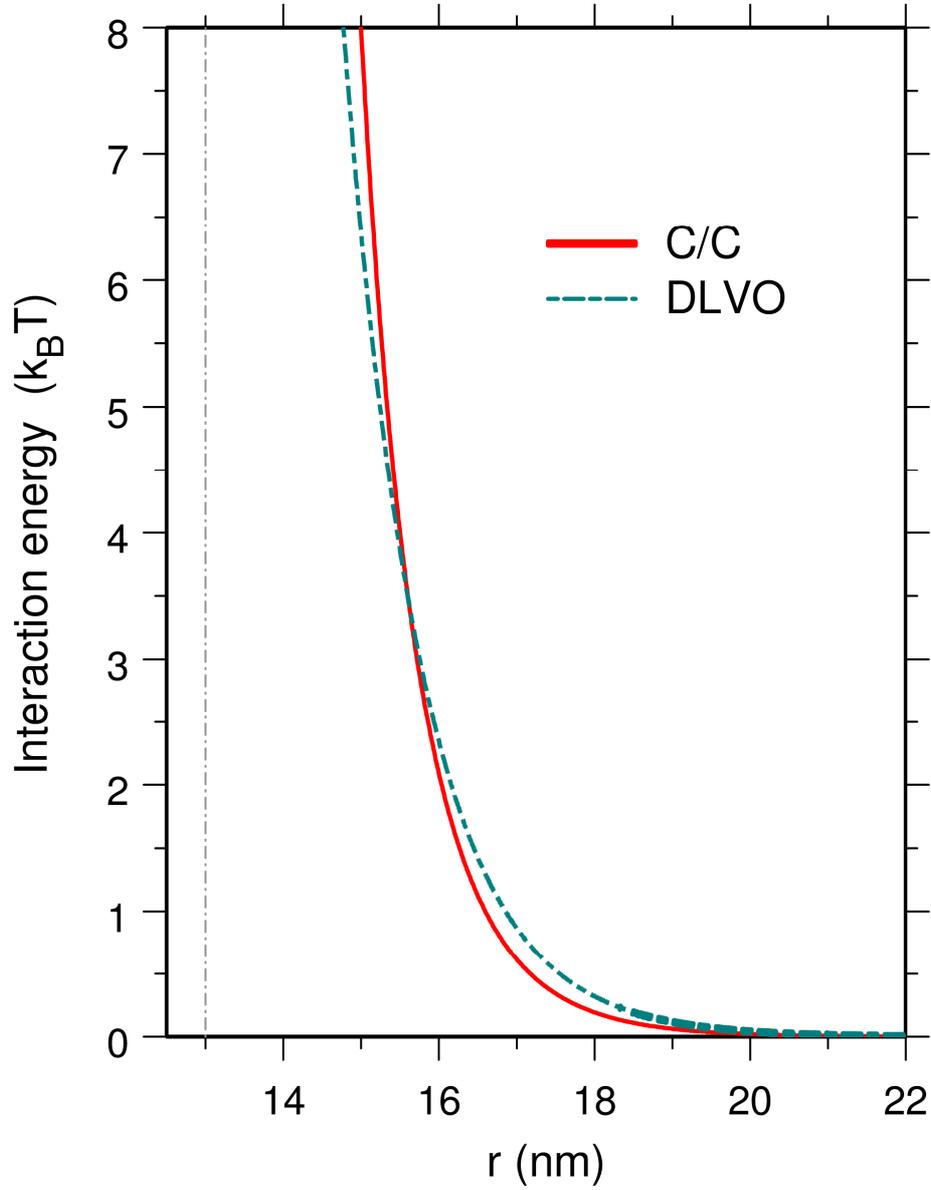

**Figure 2** : Plot of $e_C^2 \, H(r-2b)$, which describes crowder-crowder repulsion (red solid line), and the alternative DLVO potential of Eq. (5) (cyan dashed line), as a function of $r$, for $e_C = e_{DNA}$, $b = 6.5$ nm, and $e_M = -210\,\bar{e}$. The vertical dash-dotted line is located at $r = 2b$.

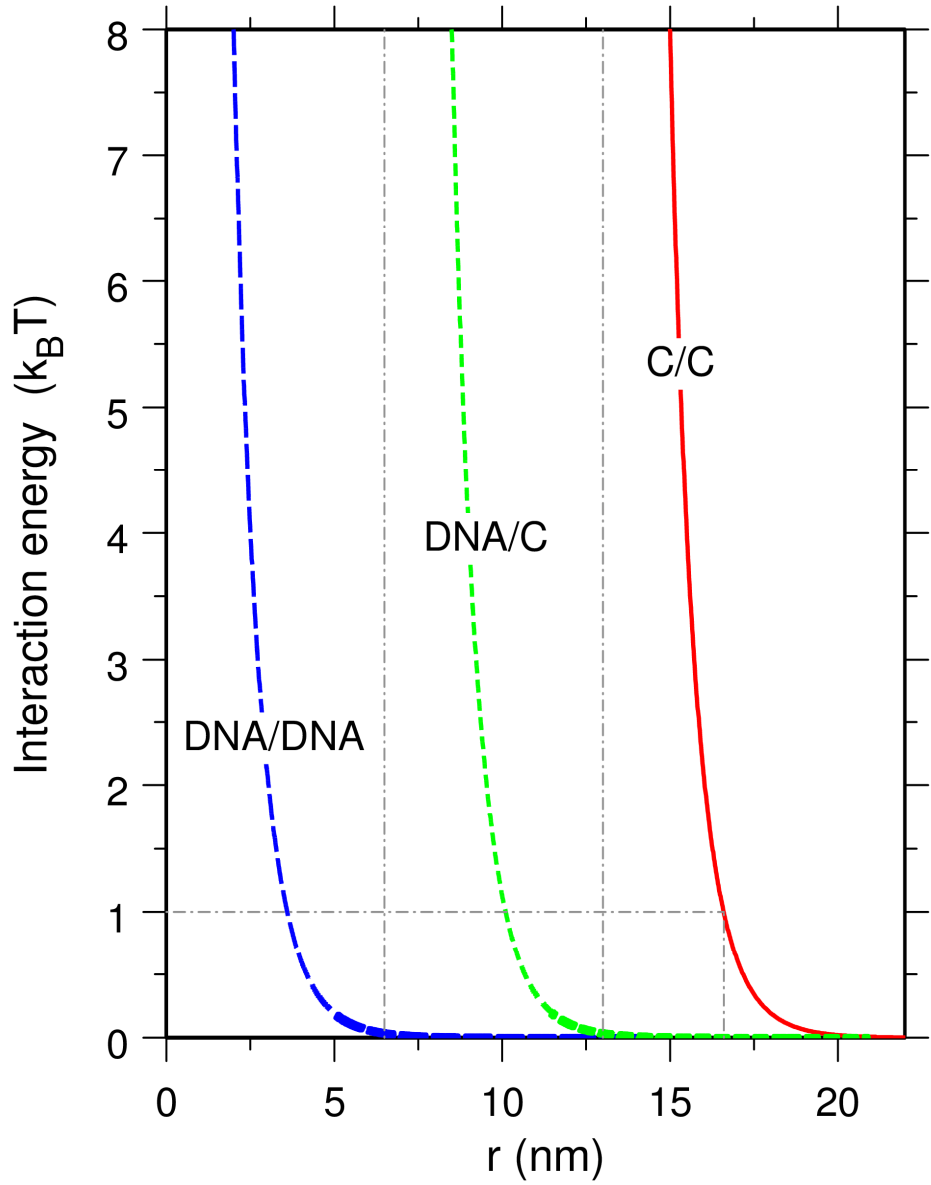

**Figure 3** : Plot of $e_{DNA}^2 H(r)$, which describes DNA-DNA repulsion (blue long-dashed line), $e_{DNA} e_C H(r-b)$, which describes DNA-crowder repulsion (green short-dashed line), and $e_C^2 H(r-2b)$, which describes crowder-crowder repulsion (red solid line), as a function of $r$, for $e_C = e_{DNA}$ and $b = 6.5$ nm. The two vertical dash-dotted lines are located at $r = b$ and $r = 2b$. The horizontal dash-dotted line located at $1 k_B T$ indicates that thermal energy corresponds to the repulsion of two spheres, which centers are separated by 16.6 nm (see text).

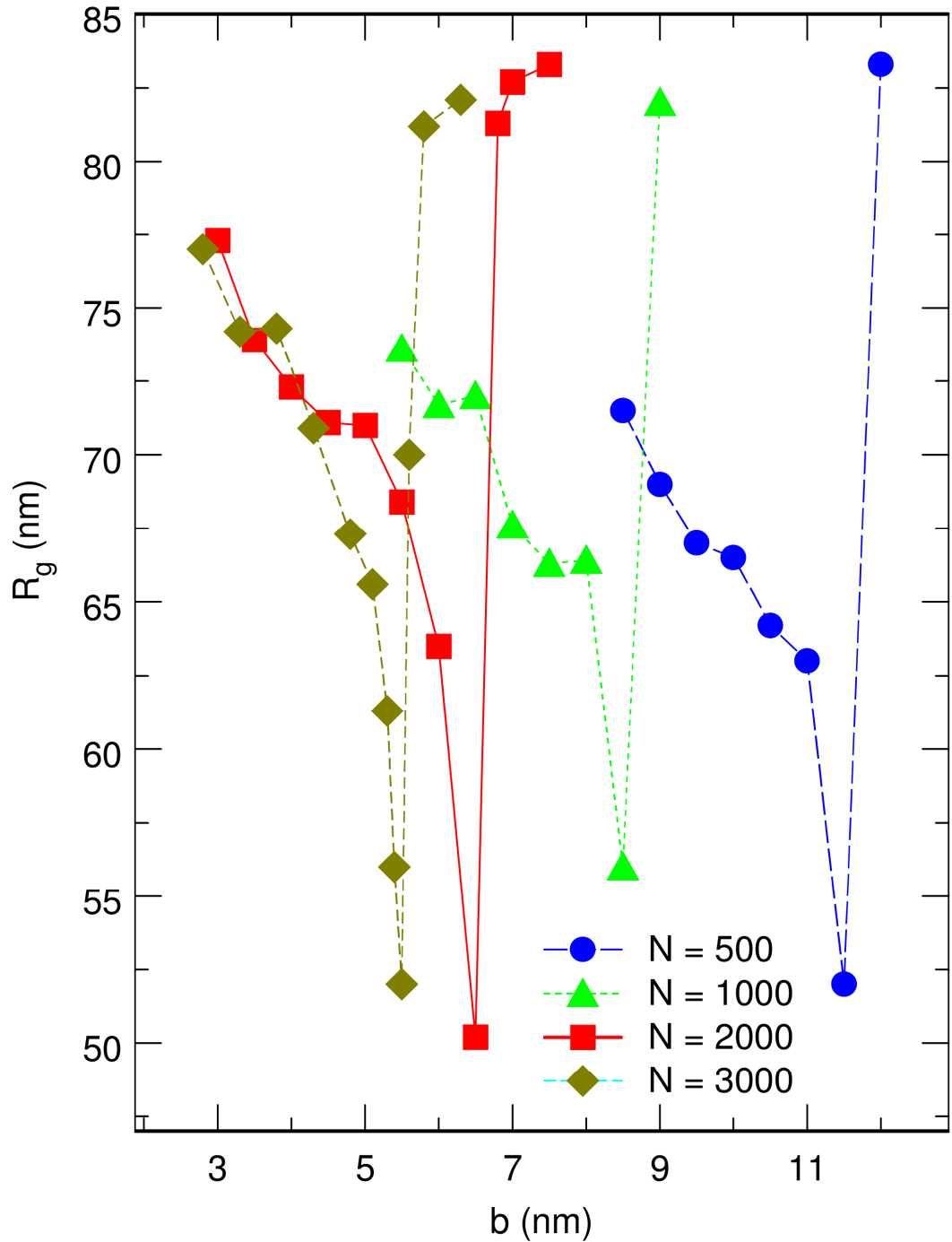

**Figure 4** : Plot of $R_g$, the mean radius of gyration of the equilibrated DNA chain, as a function of $b$, the radius of the crowding spheres, for $N$=500 (circles), 1000 (triangles), 2000 (squares) and 3000 (lozenges) crowding spheres. Both $R_g$ and $b$ are expressed in nm. Simulations were performed with $e_C = e_{DNA}$.

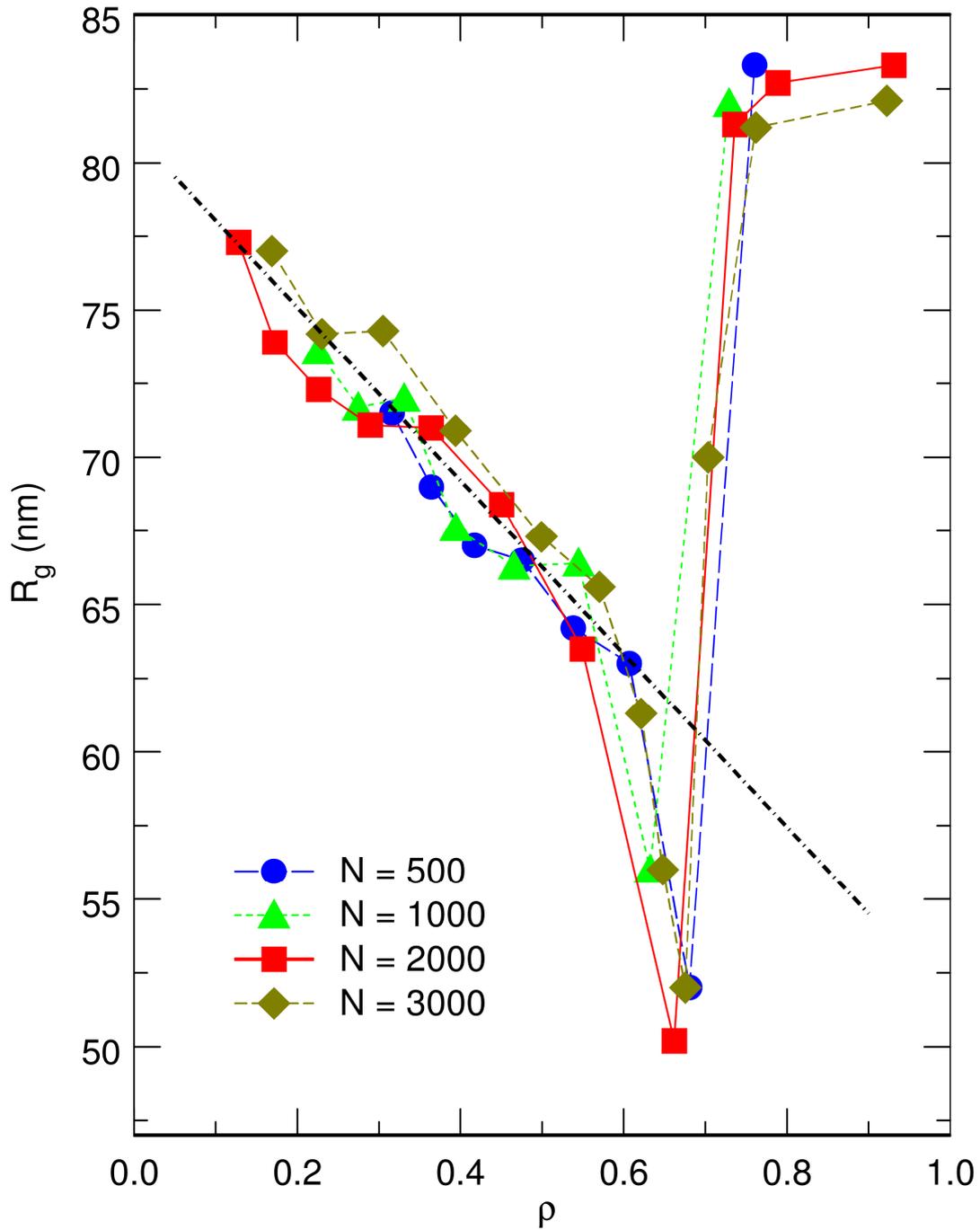

**Figure 5** : Plot of $R_g$, the mean radius of gyration of the equilibrated DNA chain, as a function of $\rho$, the effective volume occupancy ratio of the crowding spheres, for $N$=500 (circles), 1000 (triangles), 2000 (squares) and 3000 (lozenges) crowding spheres. $R_g$ is expressed in nm. Simulations were performed with $e_C = e_{DNA}$. The black dash-dotted line is just a guide for the eyes.

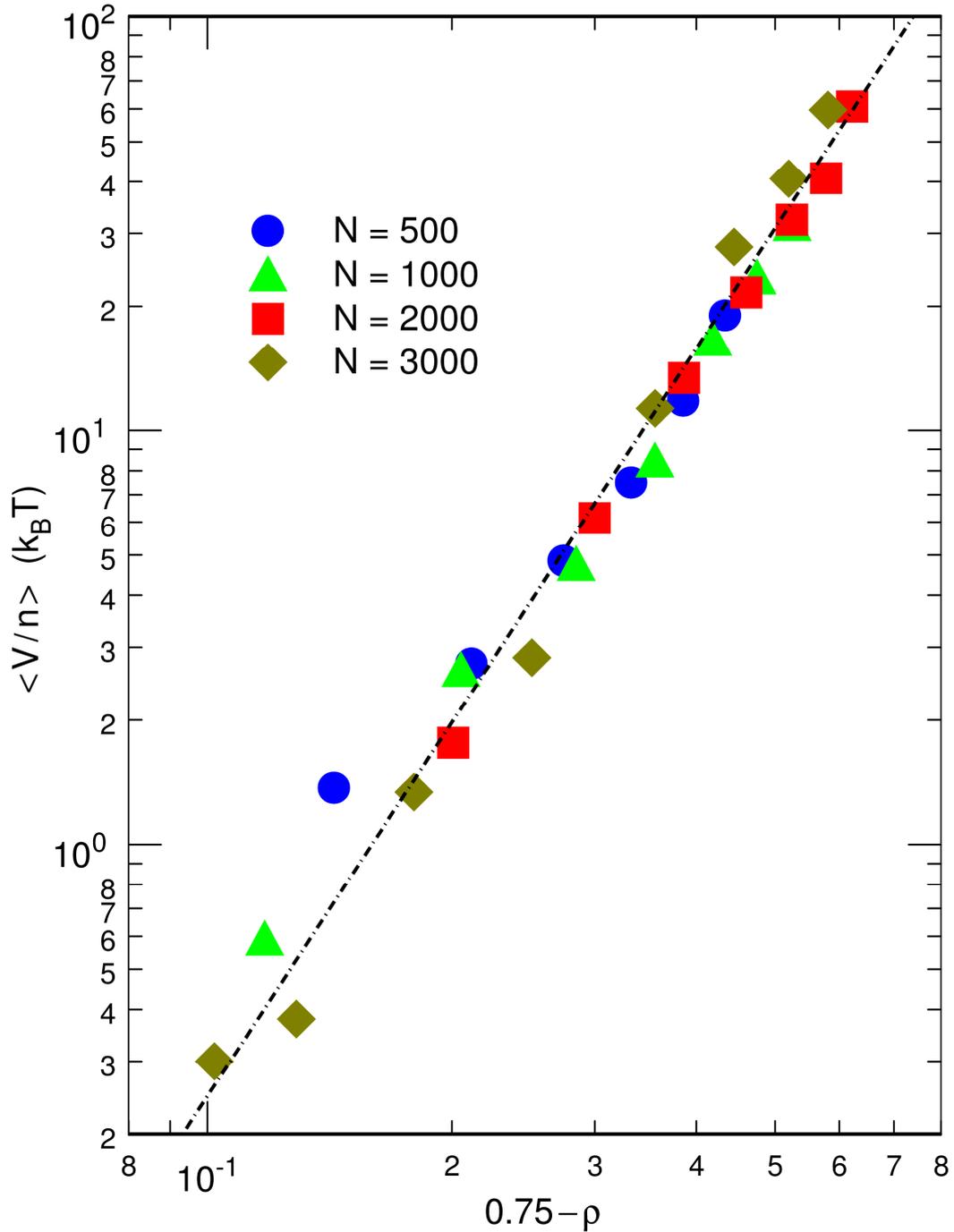

**Figure 6** : Log-log plot of the mean wall/DNA interaction energy per DNA bead as a function of $0.75-\rho$, the gap to the volume occupancy ratio at jamming, for *N*=500 (circles), 1000 (triangles), 2000 (squares) and 3000 (lozenges) crowding spheres. The mean interaction energy is expressed in units of $k_BT$. Simulations were performed with $e_C = e_{DNA}$. The black dash-dotted line is a guide for the eyes that helps visualize cubic evolution.

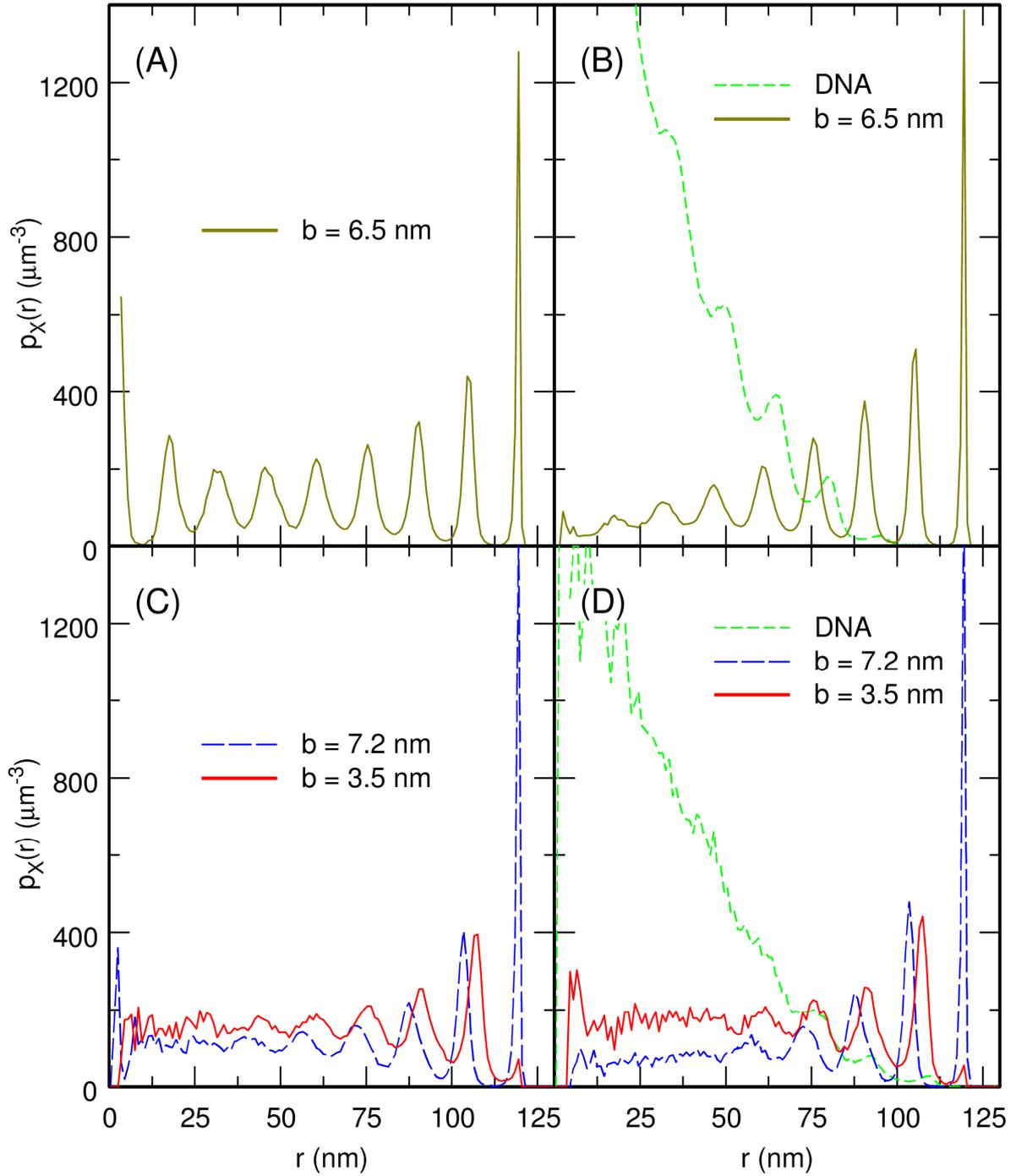

**Figure 7**: Plot of the mean density distributions of DNA beads and crowding spheres, $p_X(r)$, for equilibrated systems without DNA (vignettes A and C) or with DNA (vignettes B and D), and with 2000 crowding spheres with radius $b$=6.5 nm (vignettes A and B) or 1400 spheres with radius $b$=7.2 nm and 600 spheres with radius $b$=3.5 nm (vignettes C and D).